\documentclass[12pt]{article}
\pdfoutput=1
\usepackage{epsf,amsfonts,amssymb,epsfig,amsmath,mathtools,graphics,slashed}
\usepackage{hyperref,hep,graphicx,subfig}
\usepackage{rotating}

\newcommand{\nn}{\notag \\}

\makeatletter

\@addtoreset{equation}{section}
\makeatother

\begin{document}

\begin{titlepage}

\vfill

\begin{flushright}
Imperial/TP/2017/JG/02\\
DCPT-17/07
\end{flushright}

\vfill

\begin{center}
   \baselineskip=16pt
   {\Large\bf Holographic DC conductivity and\\
   Onsager relations}
  \vskip 1.5cm
  \vskip 1.5cm
Aristomenis Donos$^1$, Jerome P. Gauntlett$^2$\\ Tom Griffin$^2$, Nakarin Lohitsiri$^3$ and Luis Melgar$^2$\\
     \vskip .6cm
     \begin{small}
      \textit{$^1$Centre for Particle Theory and Department of Mathematical Sciences\\Durham University, Durham, DH1 3LE, U.K.}
        \end{small}\\    
         \begin{small}\vskip .6cm
      \textit{$^2$Blackett Laboratory, 
  Imperial College\\ London, SW7 2AZ, U.K.}
        \end{small}\\
         \begin{small}\vskip .6cm
      \textit{$^3$ Department of Applied Mathematics and Theoretical Physics\\ University of Cambridge,
Cambridge, CB3 OBA, UK}
        \end{small}\\
        \end{center}
     \vskip .6cm
\vfill

\begin{center}
\textbf{Abstract}
\end{center}
\begin{quote}
Within holography the DC conductivity can be obtained by solving a system of Stokes
equations for an auxiliary fluid living on the black hole horizon. We show that these
equations can be derived from a novel variational principle involving a functional that depends on the
fluid variables of interest as well as the time reversed quantities. This leads to simple derivation of the
Onsager relations for the conductivity. We also obtain the relevant Stokes equations for bulk
theories of gravity in four dimensions including a $\vartheta F\wedge F$ term in the Lagrangian,
where $\vartheta$ is a function of dynamical scalar fields. We discuss various realisations of
the anomalous Hall conductivity that this term induces 
and also solve the Stokes equations for holographic lattices which break translations in
one spatial dimension.

\end{quote}

\vfill

\end{titlepage}

\newcommand{\be}{\begin{equation}} \newcommand{\ee}{\end{equation}}
\newcommand{\bea}{\begin{eqnarray}} \newcommand{\eea}{\end{eqnarray}}

\setcounter{equation}{0}
\section{Introduction}

Holographic lattices
provide a general arena for studying thermoelectric DC transport in the context of holography 
\cite{Horowitz:2012ky}.
These are stationary black hole spacetimes with asymptotic boundary conditions that explicitly
break translation invariance of the dual field theory. This breaking of translation invariance provides a mechanism for
momentum to dissipate in the field theory and hence leads to finite results for the DC thermoelectric conductivity matrix.

It has recently been understood that when the DC conductivity matrix is finite it can be obtained, universally, by solving a system of time
independent, forced, linearised Navier-Stokes equations for an incompressible fluid on the black hole horizon \cite{Donos:2015gia,Banks:2015wha,Donos:2015bxe,Donos:2017oym}. 
While these hydrostatic fluid equations, which we refer to as Stokes equations, can be used to extract the DC conductivities, in general they are only indirectly related to physical quantities in the dual field theory. However, in the special situation of
the hydrodynamic limit of the holographic lattice, in which the temperature is the highest scale, one can
show that the horizon fluid corresponds to the hydrodynamical fluid of the dual field theory in the presence of external DC electric fields and thermal gradients \cite{Banks:2016krz}, 
thus making contact with the work on fluid-gravity \cite{Bhattacharyya:2008jc}.

This general formalism for obtaining the DC conductivity was first derived for holographic lattices 
preserving time reversal invariance \cite{Donos:2015gia,Banks:2015wha} and then extended to holographic lattices in which 
it can also be broken \cite{Donos:2015bxe}. In particular, the analysis of \cite{Donos:2015bxe} allowed for the possibility that
the dual field theory in thermal equilibrium is placed in an external magnetic field. It also covered the possibility of
having local electric and thermal magnetisation currents that are
either driven by external sources or are generated spontaneously.

On general grounds, the thermoelectric conductivity should satisfy a set of Onsager relations, which relate the 
conductivity in a given set-up to the conductivity in a time reversed setup. For the case of 
holographic lattices which preserve time reversal invariance this was shown\footnote{Onsager relations were also 
demonstrated for the electrical conductivity in the context of a hydrodynamic analysis of transport developed
in \cite{Lucas:2015lna}. This paper also
emphasised the utility of using variational principles to extract bounds on conductivities.}
in \cite{Banks:2015wha}. This was achieved by showing that the relevant
Stokes equations can be obtained from a variational principle for a functional of the fluid variables.
For holographic lattices in which the time reversal invariance is broken, a derivation of the Stokes equations
from a similar variational principle did not seem possible and hence a derivation of the Onsager relations was obscure.

In this paper we resolve this puzzle. We show that, in general, the Stokes equations of \cite{Donos:2015bxe} can be derived
from a variational principle that involves a functional that depends on the fluid variables in the background of interest and, in addition, the
time reversed quantities\footnote{We thank the referee for pointing out that a similar 
construction has been used for a dissipative wave equation in e.g. page 256 of \cite{morse1968theoretical}.}. This novel generalised variational principle then leads to a simple derivation of the Onsager relations.

Resolving this issue was the original motivation for this work. We have also taken the opportunity of extending 
the analysis of \cite{Donos:2015gia,Banks:2015wha,Donos:2015bxe}
to cover holographic theories in four spacetime dimensions which include the term $S_\vartheta\sim\int \vartheta F\wedge F$ in the action,
where $\vartheta$ is a function of dynamical scalar fields, and $F$ is the field strength of a bulk gauge field. 
A case of particular interest is when $\vartheta$
is taken to be an odd function of the pseudoscalar fields so that the theory preserves
time-reversal invariance. Indeed such terms arise very naturally in the context of consistent truncations arising from string and M-theory. Such terms are also interesting since they allow for simple holographic constructions in which the time-reversal invariance is broken spontaneously. For example the model \cite{Gauntlett:2009bh} with a single pseudoscalar
has been used to study phases that spontaneously break translations and also time reversal invariance at finite charge density \cite{Donos:2011bh,Rozali:2012es,Donos:2013wia,Withers:2013loa,Donos:2015eew}. Recently, these constructions have been generalised to holographic lattices
by including a spatially modulated chemical potential \cite{Andrade:2017leb}. Here we will outline another construction of a holographic lattice in
which time reversal invariance can be broken spontaneously at zero charge density.

The term $S_\vartheta$ gives some interesting modifications to the Stokes equations at the black hole horizon. We will see that the equations can be written naturally in a form in which only the constitutive relations for the electric current of the auxiliary fluid at the horizon
are modified and this gives rise to interesting Hall conductivity effects\footnote{In the early pioneering paper \cite{Iqbal:2008by} a term $S_\vartheta$ was considered in a simplified setting,
but some incorrect conclusions were made concerning the DC conductivity.}.
We also show that the Stokes equations on the horizon have an S-duality symmetry even if the full $D=4$ theory does not. 

The plan of the rest of the paper is as follows. In section \ref{frame} we discuss the theories of gravity that we are considering. We also briefly summarise some general features of \cite{Donos:2015gia,Banks:2015wha,Donos:2015bxe} in order to make this 
paper self contained. The Stokes equations are presented in section \ref{nsder}, where we discuss some of their properties, including a derivation of the Onsager relations. In section \ref{examples} we will discuss the impact of the term
$S_\vartheta$ on the DC conductivity in various different contexts. 
We will explain how it gives rise to what might be called
quantum critical Hall conductivity, by which we mean a contribution to the Hall conductivity for a quantum critical point that is independent of any mechanism
for momentum dissipation. In particular, this arises with vanishing applied magnetic field and zero charge density and
is thus a realisation of anomalous Hall conductivity (for a review see \cite{2010RvMP...82.1539N}).
We will also present the explicit solution of the Stokes equations, in terms of
horizon data, in the special case of holographic lattices that just depend on one of the spatial directions, thus extending the results 
presented in \cite{Donos:2015gia,Banks:2015wha,Donos:2015bxe}. We briefly conclude in section 5.

\section{Framework}\label{frame}
We will focus on the following action in four spacetime dimensions that couples the metric $g_{\mu\nu}$ to a gauge field
$A_\mu$, with field strength $F_{\mu\nu}$ and a pseudoscalar field $\phi$:
\begin{align}\label{actover}
S=\int d^4x\sqrt{-g}\left(R-V(\phi)-\frac{1}{4}Z(\phi)F^2-\frac{1}{2}(\partial \phi)^2\right)+S_\vartheta\,,
\end{align}
where
\begin{align}\label{stheta}
S_\vartheta
=\frac{1}{2}\int \vartheta(\phi) F\wedge F
=-\frac{1}{8}\int d^4x\sqrt{-g}\vartheta(\phi) \epsilon^{\mu\nu\sigma\rho}{F}_{\mu\nu}{F}_{\sigma\rho}\,,
\end{align}
with $\epsilon_{trxy}=\sqrt{-g}$. 
The action is invariant under the time reversal transformation
$t\to -t$, $A_t\to A_t$, $(A_r,A_x,A_y)\to -(A_r,A_x,A_y)$
with $\phi\to -\phi$ provided that $V(\phi)$, $Z(\phi)$ are even functions and $\vartheta(\phi)$ is an odd function of the pseudoscalar $\phi$. It is then also invariant under the parity
transformation $x\to -x$, $A_x\to -A_x$, $(A_t, A_r,A_y)\to (A_t,A_r,A_y)$
with $\phi\to -\phi$. 
We will focus on such theories in the sequel, but we note that most of our analysis is valid for arbitrary choices of the functions appearing in \eqref{actover}. The generalisation of \eqref{actover}
to include additional pseudoscalars all of which are odd under time reversal (and parity), as well as scalars that are even under time reversal (and parity), is immediate. 
We also note $A_\mu\to -A_\mu$ is a discrete symmetry of the action, associated with charge conjugation in the dual field theory.  
We will also assume that $V(\phi)$ is such that there is a unit radius $AdS_4$ vacuum solution with $A=\phi=0$.

It is also interesting to point out that for certain choices of $Z,\vartheta$ the equations of motion are invariant under $SL(2,R)$ symmetries, generated by 
$S$-duality as well as shifts of $\vartheta$ by a constant.
The $S$-duality transformation is $\phi\to -\phi$ with
\begin{align}\label{sdualft}
F_{\mu\nu}\to {Z}*F_{\mu\nu}-\vartheta F_{\mu\nu}\,,
\end{align}
where $*F_{\mu\nu}=\tfrac{1}{2}\epsilon_{\mu\nu\rho\sigma}F^{\rho\sigma}$,
provided that $Z\to\frac{Z}{Z^2+\vartheta^2}$ and
$\vartheta\to-\frac{\vartheta}{Z^2+\vartheta^2}$. A specific example is
$Z=1/\cosh(\sqrt{3}\phi)$, $\vartheta=\tanh(\sqrt{3}\phi)$
which arises as a consistent truncation of D=11 supergravity on an arbitrary seven dimensional Sasaki-Einstein space \cite{Gauntlett:2009bh}, with a skew-whiffed $AdS_4\times SE_7$ vacuum. In this truncation the pseudoscalar field is dual to a
relevant operator with\footnote{When $SE_7=S^7$ supersymmetry implies that $\Delta=2$. For other $SE_7$ the $AdS_4\times SE_7$
vacua are not supersymmetric but they are perturbatively stable \cite{Duff:1984sv}. Since the dual CFTs are not known a quantisation with
$\Delta=1$ may also be possible.} conformal scaling dimension $\Delta=2$. For the case of $SE_7=S^7$ the pseudoscalar is part of the $U(1)^4$ invariant sector of the maximally supersymmetric gauged supergravity theory \cite{Cvetic:1999xp} and hence, in particular, survives the quotient of the $S^7$ giving rise to ABJM theory.

\subsection{Holographic lattices}
We are interested in studying holographic lattices. These are defined to be 
stationary black hole solutions that approach $AdS_4$ in the UV with deformations that are associated with  
explicit breaking of translation invariance of the dual CFT. The black hole horizon is assumed to be a Killing horizon with non-zero temperature.
To simplify the exposition, we will assume that there is a single black hole, with planar topology, 
and that we can choose coordinates 
$(t,r,x,y)$, with $\partial_t$ the stationary Killing vector, which are globally defined outside the horizon.  
More general setups can also be studied, as explained in \cite{Banks:2015wha,Donos:2015bxe}.

We then demand that as we approach $AdS_4$ at $r\to \infty$, we have
\begin{align}\label{asmet}
ds^2&\to r^{-2}dr^2+r^2\left[g^{(\infty)}_{tt}dt^2+g^{(\infty)}_{ij}dx^idx^j+2g_{ti}^{(\infty)}dtdx^i\right]\,,\nn
A&\to A^{(\infty)}_t dt+A^{(\infty)}_i dx^i\,, \qquad
\phi\to r^{\Delta-3}\phi^{(\infty)}\,,
\end{align}
where $g^{(\infty)}_{tt}$ etc. are functions of the spatial coordinates, $x^i=(x,y)$, only, and give rise to breaking the translation
invariance of the dual field theory. These will be taken to be periodic in the spatial coordinates $x^i$ with the exception\footnote{In the special case that we have massless axion fields with $\Delta=3$, we can also consider the axion to be linear in the spatial coordinates.}
that we allow for a constant magnetic field with strength $B$ by writing $A^{(\infty)}_i =-\frac{1}{2}B\epsilon(ij)x^j+\hat A^{(\infty)}_i$
with periodic $\hat A^{(\infty)}_i$ and $\epsilon(xy)=1$.

Note that the source terms $g_{ti}^{(\infty)}$, $A^{(\infty)}_i$ and $\phi^{(\infty)}$ 
are all odd under the time reversal transformation that we discussed above and 
explicitly break time reversal invariance. 
It is important to note that if these terms vanish, we can still have holographic lattices that spontaneously break time reversal invariance and we will return to this point later.

We assume that the black hole horizon is located at $r=0$. We introduce
a function $U(r)$ with analytic expansion $U\left(r\right)=4\pi\,Tr+\dots$ and demand that
as $r\to 0$ we have
\begin{align}\label{defhqs}
g_{tt}(r,x)&=-U(G^{(0)}(x)+...)\,,\qquad
g_{rr}(r,x)=U^{-1}(G^{(0)}(x)+...)\,,\nn
g_{tr}(r,x)&=U(g_{tr}^{(0)}(x)+...)\,,\qquad
g_{ti}(r,x)=U({G^{(0)}(x)} \chi_i^{(0)}(x)+...)\,,\nn
A_t(r,x)&=U(\frac{G^{(0)}(x)}{4\pi T}A_t^{(0)}(x)+...)\,,
\end{align}
where the dots refer to higher powers in $r$ and all other quantities are, in general, non-vanishing at the horizon:
\begin{align}\label{defhqs2}
g_{ij}(r,x)&=h_{ij}^{(0)}(x)+ ...\,,\quad
g_{ir}(r,x)=g_{ir}^{(0)}(x)+...\,,\cr
A_i(r,x)&=A_i^{(0)}(x)+...\,,\,\,
A_r(r,x)=A_r^{(0)}(x)+...\,,\,\,
\phi(r,x)=\phi^{(0)}(x)+...\,.
\end{align}
Note that the function ${G^{(0)}(x)}$ can be set to unity after carrying out a coordinate transformation, if one wishes (see footnote 8 of \cite{Donos:2017oym}).
Of most importance in the sequel is the horizon data $h_{ij}^{(0)}$, $A_t^{(0)}$, $\chi_i^{(0)}$ and $\phi^{(0)}$.

\subsection{The DC perturbation}

In order to introduce suitable DC sources for the electric and heat currents we consider the following linear perturbation
of the black hole solution\footnote{Note that this corrects a sign typo in eq. (4.1) of \cite{Donos:2015bxe}.}:
\begin{align}
\delta(ds^2)&=\delta g_{\mu\nu}dx^\mu dx^\nu+2tg_{tt}\zeta_i dt dx^i+t(g_{ti}\zeta_j+g_{tj}\zeta_i)dx^i dx^j+2tg_{tr}\zeta_idr dx^i\,,\nn
\delta A&=\delta a_\mu dx^\mu-t E_i dx^i+tA_t\zeta_i dx^i\,,\label{heatpertansatz}
\end{align}
as well as $\delta\phi$. The source terms $E=E_i(x)dx^i$ and $\zeta=\zeta_i(x)dx^i$, with periodic dependence in $x^i$, 
are closed one-forms,
$dE=d\zeta=0$.
The other functions in the perturbation depend on both $r$ and periodically on $x^i$. 
In particular all
time dependence of the equations of motion is satisfied, to linear order in the perturbation. The harmonic parts of the
sources $E,\zeta$ characterise the physical external sources.

The coordinate $t$ is no longer a good
coordinate at the black hole horizon. Switching to the ingoing Eddington-Finklestein coordinate $v=t+\ln r/(4\pi T)+\dots$, regularity of the perturbation near $r=0$
places the following restrictions on $\delta g_{\mu\nu},\delta a_\mu,\delta\phi$ at the horizon,
\begin{align}
\delta g_{tt}&=U(\delta g_{tt}^{(0)}(x)+...)\,,\qquad\qquad\quad
\delta g_{rr}={U}^{-1}(\delta g_{rr}^{(0)}(x)+...)\,,\cr
\delta g_{rt}&=\delta g_{rt}^{(0)}(x)+...\,,\qquad\qquad\qquad
\quad\delta g_{ti}=\delta g_{ti}^{(0)}(x)+g_{tt}\frac{\ln r}{4\pi T}\zeta_i+...\,,\cr
\delta g_{ij}&=\delta g_{ij}^{(0)}(x)+\frac{2\ln r}{4\pi T}g_{t(i}\zeta_{j)}+...\,,\quad
\delta g_{ri}={U}^{-1}\delta g_{ti}^{(0)}(x)+\frac{\ln r}{4\pi T} g_{tr}\zeta_i+...\,,
\end{align}
where $\delta g_{tt}^{(0)}+\delta g_{rr}^{(0)}-2\delta g_{rt}^{(0)}=0$. We also have
\begin{align}
\delta a_t&=\delta a_t^{(0)}(x)+...\,,\qquad\qquad\qquad
\delta a_r={U}^{-1}(\delta a_t^{(0)}(x)+...)\,,\cr
\delta a_j&=\frac{\ln r}{4\pi T}(-E_j+A_t\zeta_j)+\delta a^{(0)}_j(x)...\,,\qquad\delta\phi=\delta\phi^{(0)}(x)+...\,.
\end{align}

\subsection{Electric and heat currents}
We begin with the electric currents which are simpler. We first define the two form
\begin{align}
{H}_{\mu\nu}\equiv Z{F}_{\mu\nu}+{\vartheta}*{F}_{\mu\nu}\,,
\end{align}
and then the equation of motion for the gauge field is $\partial_\mu (\sqrt{-g}{H}^{\mu\nu})=0$.
We define the bulk electric current density to be
\begin{align}
J^a
=\sqrt{-g}{H}^{ar}.
\end{align}
When evaluated at the $AdS$ boundary $J^t_\infty$ is the local charge density and $J^i_\infty$ is the local electric current density
of the dual field theory. In the presence of the DC perturbation we deduce that
\begin{align}\label{drJ}
\partial_r J^i&=\partial_j (\sqrt{-g}{H}^{ji})+\sqrt{-g}{H}^{ij}\zeta_j\,,\nn
\partial_i J^i&=\zeta_iJ^i\,.
\end{align}
We also have
\begin{align}\label{chgd}
\partial_r J^t=\partial_j (\sqrt{-g}H^{jt})\,.
\end{align}

We next discuss the bulk heat current density\footnote{It is also possible to follow the approach developed in \cite{Donos:2017oym}. For example, in eq. (3.4) of \cite{Donos:2017oym} one would have
$2W_{mn}=H^2w_{mn}-HA_t(Zv_{mn}+\vartheta(\bar*u)_{mn})$.}. We first define
\begin{align}
G^{\mu\nu}\equiv-2\nabla^{[\mu}k^{\nu]}-Zk^{[\mu}{F}^{\nu]\sigma}A_\sigma-\frac{1}{2}(\varphi-\theta){H}^{\mu\nu}\,,
\end{align}
where $k^\mu$ is a vector field satisfying $\nabla_\mu k^\mu=0$, $\varphi\equiv i_kA$ and $\psi\equiv i_k\mathcal{F}-d\theta$ for any globally defined function $\theta$. For our setup, we will take $k^\mu=(\partial_t)^\mu$ and $\theta=-A_t$, so that $\varphi=-\theta=A_t$ and $\psi_\nu=\partial_t A_\nu$. Note that with the DC perturbation $k^\mu=(\partial_t)^\mu$ is no longer a Killing vector, but we still have $\nabla_\mu k^\mu=0$. We have
\begin{align}
\nabla_\mu G^{\mu\nu}=&(-\nabla_\mu\zeta^\mu+V)k^\nu+dk^{\nu\rho}\zeta_\rho+\frac{1}{2}Z{F}^{\nu\mu}\psi_\mu-\frac{ZA_\sigma{L}_k({F}^{\nu\sigma})}{2}\cr
&\qquad -\frac{\partial_\lambda\vartheta}{4}\epsilon^{\lambda\sigma\tau\rho}{F}_{\tau\rho}A_\sigma k^\nu-\frac{\vartheta}{4}\epsilon^{\mu\nu\sigma\rho}{F}_{\sigma\rho}\nabla_\mu (\varphi-\theta)\,.\label{Gder}
\end{align}
We can then define the bulk heat current density as
\begin{align}
Q^i\equiv\sqrt{-g}G^{ir}.
\end{align}
When evaluated at the $AdS$ boundary we have $Q^i_\infty$ is the heat current density of the dual field theory.
After some calculation we find that in the presence of the DC perturbation we have
\begin{align}
\partial_rQ^i&=\partial_j(\sqrt{-g} G^{ji})+2\sqrt{-g}G^{ij}\zeta_j+\sqrt{-g}{H}^{ij} E_j\,,\nn
\partial_iQ^i&=2Q^i\zeta_i+J^iE_i.\label{drQ}
\end{align}

We next rewrite the bulk equations of motion using a radial Hamiltonian decomposition. Since the extra term $S_\vartheta$
is independent of the metric, the only change in the analysis of \cite{Donos:2015bxe} 
is a redefinition of the momentum conjugate to the gauge fields. Denoting the momentum conjugate to the metric and gauge-field
on the radial hypersurfaces as $\pi^{ab}$ and $\pi^a$, we still have
\begin{align}
J^a&=\pi^a\,,\nn
Q^i &=-2{\pi^i}_t-\pi^iA_t\,.
\end{align}
The latter expression confirms that
$Q^i_\infty$ is indeed the local heat current density of the dual field theory.
By evaluating the hamiltonian, momentum and Gauss law constraints on the stretched horizon, gives rise to a system of 
Stokes equations on the horizon, that will be given in the next subsection. These equations can be solved to obtain the current densities
$J^i_{(0)}$, $Q^i_{(0)}$ on the horizon. In turn, these can be used to obtain the zero modes of suitably defined 
transport currents of the dual field theory, as we next explain. 
For further discussion on transport currents see \cite{PhysRevB.55.2344,Hartnoll:2007ih,Blake:2015ina,Donos:2017oym}.

\subsection{Transport currents}
First consider the currents for the background black holes, with vanishing DC perturbation.  One can show that
the currents vanish at the black hole horizon. Hence, upon integrating \eqref{drJ},\eqref{drQ} in the radial direction 
we deduce that the current densities for the background black holes are magnetisation currents
of the form
\begin{align}\label{jayback}
J^{(B)i}_\infty=\partial_j M^{(B)ij}\,,\qquad
Q^{(B)i}_\infty=\partial_j M^{(B)ij}_T\,,
\end{align}
where 
$M^{ij}(x)$, and 
$M^{ij}_T(x)$ are given by
\begin{align}\label{emmslocal}
M^{ij}=-\int_0^\infty dr\sqrt{-g}H^{ij} \,,\qquad
M_T^{ij}=-\int_0^\infty dr\sqrt{-g}G^{ij}\,.
\end{align}
Clearly $\partial_i J^{(B)i}_\infty=\partial_i Q^{(B)i}_\infty=0$. In addition, since the integrands are periodic functions of
the spatial coordinates, it is clear that the zero modes of these currents
must vanish $\bar{{J}}^{(B)i}_\infty=\bar{{Q}}^{(B)i}_\infty=0$,
where the bar refers to the following average integral taken over a period of the spatial coordinates:
\begin{align}\label{avges}
\bar A\equiv \frac{1}{L_1L_2}\int_0^{L_1} \int_0^{L_2} d^2 x A(x)\,.
\end{align}

In the presence of the DC perturbation we can define the local transport currents of the dual field theory to be
\begin{align}\label{caljdefs}
\mathcal{J}^i&\equiv J^i_{\infty}+M^{(B)ij}\zeta_j,\nn
\mathcal{Q}^i&\equiv Q^i_{\infty}+M^{(B)ij}E_j+2M_T^{(B)ij}\zeta_j\,,
\end{align}
with $\partial_i \mathcal{J}^i=0$ and $\partial_i\mathcal{Q}^i=0$.
From (\ref{drJ}), (\ref{drQ}) we deduce that the transport current flux densities, 
relevant for the DC conductivity, are given by the horizon current flux densities:
\begin{align}\label{reneq}
\bar{\mathcal{J}}^i=\bar J^i_{(0)},\qquad
\bar{\mathcal{Q}}^i=\bar Q^i_{(0)}\,.
\end{align}

\section{Stokes equations}\label{nsder}
By evaluating the Hamiltonian, momentum and Gauss law constraints on the black hole horizon we find a closed system of time independent, forced,
linearised Navier-Stokes equations for a subset of the perturbation $(v_i, p, w)$ defined by
\begin{align}
\label{quantdef}
v_i&\equiv-\delta g_{ti}^{(0)}\,,\qquad
p\equiv-\frac{4\pi T}{G^{(0)}}\left(\delta g_{rt}^{(0)}-h^{ij}_{(0)}g_{ir}^{(0)}\delta g_{tj}^{(0)}\right)-h^{ij}_{(0)}\frac{\partial_i G^{(0)}}{G^{(0)}}\delta g_{tj}^{(0)}\,,\nn
w&\equiv\delta a_t^{(0)}\,,
\end{align}
where $h^{ij}_{(0)}$ is the inverse metric for $h_{ij}^{(0)}$.
It is also helpful to introduce the {\it local} charge density, $\rho_H$, of the black hole background evaluated at the horizon:
\begin{align}
\rho_H\equiv J^t_{(0)}=\sqrt{h^{(0)}}\left(Z^{(0)}A^{(0)}_t- \frac{1}{2}\vartheta^{(0)}\epsilon^{ij}F^{(0)}_{ij}\right)\,,
\end{align}
with $\epsilon_{ij}=\sqrt{h^{(0)}}\epsilon(ij)$ the volume form on the horizon, where $\epsilon(xy)=1$.
In general $\rho_H$ is not the same as the charge density of the dual field theory, However, because of \eqref{chgd}, the zero mode, $\rho\equiv\bar\rho_H$, 
defined by \eqref{avges}, is not renormalised in going to the holographic boundary and hence $\rho$ is the zero mode of the charge density of the dual field theory.

The resulting system of Stokes equations can be written
\begin{align}\label{navst}
&-2{}{\nabla}^j{}{\nabla}_{(i}v_{j)}+[{{\nabla}_i\phi^{(0)}\nabla}_j\phi^{(0)}-4\pi Td\chi_{ij}^{(0)}]v^j{}
-\frac{1}{\sqrt{h^{(0)}}}F^{(0)}_{ij} J^j_{(0)}
\nn &\qquad\qquad\quad\qquad\qquad\qquad\qquad
=4\pi T(\zeta_i-\frac{1}{4\pi T}\nabla_i p)
+\frac{\rho_H}{\sqrt{h^{(0)}}}(E_i+{}{\nabla}_iw)
\,,\nn
&{}\partial_iQ^i_{(0)}=0\,,\qquad\qquad {}\partial_iJ^i_{(0)}=0\,,
\end{align}
where the current densities on the horizon are given by
\begin{align}
J^i_{(0)}&=\rho_Hv^i+ \sqrt{h^{(0)}}\left(Z^{(0)}h^{ij}_{(0)}-\vartheta^{(0)}\epsilon^{ij}\right)\left(E_j+{}{\nabla}_jw+F^{(0)}_{jk}v^k\right)\,,\nn
Q^i_{(0)}&= 4\pi T\sqrt{h^{(0)}}v^i\,.
\end{align}
These expressions depend on the following horizon quantities $h^{(0)}_{ij}$, $A^{(0)}_t$, $Z^{(0)}\equiv Z(\phi^{(0)})$, 
$\vartheta^{(0)}\equiv\vartheta(\phi^{(0)})$ and $\nabla$ is the covariant derivative with respect to $h^{(0)}_{ij}$.

By solving these equations on the horizon, we obtain expressions for the local currents on the horizon $J^i_{(0)}$, $Q^i_{(0)}$ as a function 
of the applied DC source parametrised by $E_i,\zeta_i$. Via \eqref{reneq} we can then obtain
expressions for the DC transport fluxes $\bar{\mathcal{J}}^i$, $\bar{\mathcal{Q}}$ as functions of the applied DC source, and hence
the DC conductivity.

\subsection{Uniqueness and positivity of conductivity}
It is straightforward to show that the Stokes equations imply
\begin{align}\label{pos}
\int d^2x\sqrt{h^{(0)}}\Big(2{\nabla}^{(i}v^{k)}{\nabla}_{(i}v_{k)}+[{\nabla}_j\phi^{(0)}v^j]^2+&Z^{(0)}|E_i+{\nabla}_iw+{F}^{(0)}_{ij}v^j|^2\Big)\cr
&
=\int d^2x\left( Q^i_{(0)}\zeta_i+J^i_{(0)}E_i\right)\,.
\end{align}
To examine the issue of uniqueness of solutions we set $E_i=\zeta_i=0$. From \eqref{pos} we immediately obtain
${\nabla}_{(i}v_{j)}=0$, $v^j{\nabla}_j\phi^{(0)}=0$ and ${\nabla}_iw+v^j{F}^{(0)}_{ij}=0$. Then from \eqref{navst}
we deduce ${\nabla}_ip+4\pi Tv^jd{\chi}_{ji}^{(0)}=0$ and 
$Z^{(0)}v^i{}{\nabla}_iA_t^{(0)}-\nabla_i\vartheta^{(0)}\epsilon^{ij}\nabla_jw =0$.
In particular, we see that solutions to the source free equations require that $v^i$ is a Killing vector that preserves
$\phi^{(0)}$, $w$, $p$ and that the Lie-derivative of $A_t^{(0)}$ is a specific gauge-transformation. 
For a given background, if no such $v^i, p, w$ satisfying these conditions exist, then any solution to the sourced equations will be unique.

The exact parts of the closed one-forms $E,\zeta$ do not contribute to the currents on the horizon and hence to calculate the DC conductivities
we can take $E,\zeta$ to be harmonic. Taking the components $E_i,\zeta_i$ to be constants\footnote{For a more general discussion see section 3.7 of \cite{Banks:2015wha}.}, we see that if we take the integrals in \eqref{pos} over a period and divide by $L_1L_2$,
then the right hand side is just  $\bar{\mathcal{Q}}^i\zeta_i+\bar{\mathcal{J}}^iE_i\equiv E_i\sigma^{ij}E^j+E_iT\alpha^{ij}\zeta_j+\zeta_iT\bar\alpha^{ij}E_j+\zeta_iT\bar\kappa^{ij}\zeta_j$. The positivity of the left hand side of \eqref{pos} then shows that the full thermoelectric conductivity matrix is positive semi-definite.

\subsection{Lagrangian Formulation and Onsager Relations}
We now derive Onsager relations relating the DC conductivities for a given holographic lattice to
one that is associated with deformations obtained via time reversal. In the bulk, the relevant time reversal symmetry
is obtained by taking $t\rightarrow-t$, $\phi\to -\phi$ and $A\rightarrow -A$ for the bulk gauge field.
Given a background black hole solution, this implies
that the horizon quantities of the time reversed background solution can be obtained by the transformations
${\chi}^{(0)}_{i}\rightarrow -{\chi}^{(0)}_{i}$, ${F}^{(0)}_{ij}\rightarrow -{F}^{(0)}_{ij}$ and $\vartheta^{(0)}\rightarrow-\vartheta^{(0)}$ and
we notice that $\rho_H\to \rho_H$.

We next consider the DC perturbation on the horizon that is associated with the time reversed solution. 
We will use the notation $\tilde v^i,\tilde p,\tilde w$ for the perturbation of the time reversed background that survives at the horizon, and these will be determined in terms of the
DC source which we denote $\tilde E_i,\tilde\zeta_i$ (independent of $E_i,\zeta_i$).
The currents on the horizon for the time reversed solution are then given by
\begin{align}
\tilde J^i_{(0)}&=\rho_H\tilde v^i+ \sqrt{h^{(0)}}\left(Z^{(0)}h^{ij}_{(0)}+\vartheta^{(0)}\epsilon^{ij}\right)\left(\tilde E_j+{}{\nabla}_j\tilde w-F^{(0)}_{jk}\tilde v^k\right)\,,\nn
\tilde Q^i_{(0)}&= 4\pi T\sqrt{h^{(0)}}\tilde v^i\,,
\end{align}
and they can be obtained by solving the following Stokes equations:
\begin{align}\label{navst2}
&-2{}{\nabla}^j{}{\nabla}_{(i}\tilde v_{j)}+[{{\nabla}_i\phi^{(0)}\nabla}_j\phi^{(0)}+4\pi Td\chi_{ij}^{(0)}]\tilde v^j{}
+\frac{1}{\sqrt{h^{(0)}}}F^{(0)}_{ij} \tilde J^j_{(0)}
\nn &\qquad\qquad\quad\qquad\qquad\qquad\qquad
=4\pi T(\tilde \zeta_i-\frac{1}{4\pi T}\nabla_i \tilde p)
+\frac{\rho_H}{\sqrt{h}}(\tilde E_i+{}{\nabla}_i\tilde w)
\,,\nn
&{}\partial_i\tilde Q^i_{(0)}=0\,,\qquad\qquad {}\partial_i\tilde J^i_{(0)}=0\,.
\end{align}

Remarkably, the equations for the background \eqref{navst} and the time-reversed background \eqref{navst2} can both be obtained from the
following Lagrangian:
\begin{align}\label{lag}
L=&\int d^{2}x\,\sqrt{h^{(0)}} \Bigg[-{\nabla}^{\left( i \right. }\tilde{v}^{\left. j\right)} 
{\nabla}_{\left( i \right. }v_{\left. j\right)}
+\frac{1}{2}\tilde v^i v^j( -\nabla_i\phi^{(0)}\nabla_j\phi^{(0)}+  4\pi Td{\chi}_{ij}^{(0)}+\frac{\rho_H}{\sqrt{h^{(0)}}}F^{(0)}_{ij})\nn
&+\frac{1}{2}\tilde v^i\Big(  4\pi T\zeta_i -{\nabla}_{i}p +\frac{\rho_H}{\sqrt{h^{(0)}}} ({E}_i+{}{\nabla}_i{w})  \Big)\nn
&+\frac{1}{2} v^i\left(  4\pi T\tilde \zeta_i -{\nabla}_{i}\tilde p +\frac{\rho_H}{\sqrt{h^{(0)}}}({\tilde E}_i+{}{\nabla}_i{\tilde w})  \right)\nn
&+\frac{1}{2}({}{\tilde E_{i}+\nabla}_{i}\tilde w-{F}^{(0)}_{ik}\tilde v^k)(Z^{(0)}h_{(0)}^{ij}-\vartheta^{(0)}\epsilon^{ij})({}{E_{j}+\nabla}_{j} w +F^{(0)}_{jl} v^l)
\Bigg]\,.
\end{align}

Variations with respect to $(\tilde{v}^i,\tilde{p},\tilde{w})$ yield the system \eqref{navst} and variations with respect to  
$(v^i,p,w)$
yield \eqref{navst2}. Furthermore,
we have:
\bea
J^{i}_{(0)}=2\,\frac{\delta L}{\delta \tilde{E}_{i}},\quad Q^{i}_{(0)}=2\,\frac{\delta L}{\delta \tilde{\zeta}_{i}}\,,\qquad\qquad
\tilde{J}^{i}_{(0)}=2\,\frac{\delta L}{\delta E_{i}},\quad \tilde{Q}^{i}_{(0)}=2\,\frac{\delta L}{\delta \zeta_{i}}\,.
\eea
By taking an additional derivative with respect to the sources and commuting partial derivatives we immediately obtain the
Onsager relations for the local current densities on the horizon, relating quantities on the background geometry with those in
the time-reversed geometry. On-shell, we have
\begin{align}
\frac{\delta J^{i}_{(0)}}{\delta {E}_{j}}=\frac{\delta \tilde J^{j}_{(0)}}{\delta \tilde{E}_{i}}\,,\qquad
\frac{\delta J^{i}_{(0)}}{\delta {\zeta}_{j}}=\frac{\delta \tilde Q^{j}_{(0)}}{\delta \tilde{E}_{i}}\,,\qquad
\frac{\delta Q^{i}_{(0)}}{\delta {E}_{j}}=\frac{\delta \tilde J^{j}_{(0)}}{\delta \tilde{\zeta}_{i}}\,,\qquad
\frac{\delta Q^{i}_{(0)}}{\delta {\zeta}_{j}}=\frac{\delta \tilde Q^{j}_{(0)}}{\delta \tilde{\zeta}_{i}}\,.
\end{align}
In these expressions the left hand side depends on the horizon data $(F^{(0)}_{ij},\chi^{(0)}_i,\vartheta^{(0)})$ while the right hand side depends
on the horizon data $(-F^{(0)}_{ij},-\chi^{(0)}_i,-\vartheta^{(0)})$ which arises in the time reversed background.

The transport currents of the time reversed configuration can be written
\begin{align}\label{caljdefs2}
\tilde{\mathcal{J}}^i&\equiv \tilde J^i_{\infty}+\tilde M^{(B)ij}\tilde \zeta_j,\nn
\tilde{\mathcal{Q}}^i&\equiv \tilde Q^i_{\infty}+\tilde M^{(B)ij}\tilde E_j+2\tilde M_T^{(B)ij}\tilde\zeta_j\,,
\end{align}
and we have $\tilde M^{(B)ij}=-M^{(B)ij}$, $\tilde M_T^{(B)ij}=- M_T^{(B)ij}$. By solving the Stokes equations \eqref{navst2}
we can obtain the current fluxes at the horizon which are identical to fluxes of the transport currents in the dual CFT 
and hence we immediately obtain the Onsager relations of the dual CFT. 
Specifically, if we let $S$ denote the UV data specifying the holographic lattice and $S^t$ as the time-reversed data, with
$(S^t)^t=S$, we have
\begin{align}\label{simpon}
\sigma^T(S)=\sigma(S^t),\qquad \alpha^T(S)=\bar{\alpha}(S^t),\qquad \bar{\kappa}^T(S)=\bar{\kappa}(S^t)\,.
\end{align}

Note that we have presented the Lagrangian \eqref{lag} and the Onsager relations for the specific bulk theory in
four spacetime dimensions \eqref{actover}. However, it is straightforward to obtain them for other holographic theories. For example,
if we consider the theory with Lagrangian \eqref{actover} and $S_\vartheta=0$ but in any spacetime dimension, then the
resulting Stokes equations can be obtained by varying the integrand in \eqref{lag} after setting $\vartheta=0$.
Finally, it is also worth pointing out that on-shell we have
\begin{align}
L_{OS}=\frac{1}{2}\int d^{2}x \Big[\tilde E_i J^i_{(0)} + \tilde \zeta_i Q^i_{(0)} \Big]=\frac{1}{2}\int d^{2}x \Big[ E_i \tilde J^i_{(0)}  +\zeta_i \tilde Q^i_{(0)} \Big]\,.
\end{align}

\subsection{S-duality}\label{sdualsec}
It is also interesting to note that the system of Stokes equations \eqref{navst} is invariant under S-duality transformations of the horizon quantities.
We emphasise that this is independent of whether or not the bulk theory is invariant under the $S$-duality transformation discussed near
\eqref{sdualft}. We first define the local magnetic field on the horizon, $B_H$, via
\begin{align}
B_H\equiv \frac{1}{2}\sqrt{h^{(0)}}\epsilon^{ij}F^{(0)}_{ij}\,.
\end{align}
Notice that this is independent of the metric on the horizon. We also point out that the zero mode $B\equiv \bar{B}_H$, being a 
topological charge, is not renormalised in going to the boundary and is in fact the total magnetic field of the dual field theory.
 The S-duality transformations can then be written
\begin{align}\label{stran}
B_H&\to \rho_H,\qquad \qquad\qquad\quad\rho_H\to-B_H\,,\nn
Z^{(0)}&\to\frac{Z^{(0)}}{Z_{(0)}^2+\vartheta_{(0)}^2}\,,\qquad\quad
\vartheta^{(0)}\to-\frac{\vartheta^{(0)}}{Z_{(0)}^2+\vartheta_{(0)}^2}\,\nn
(E_i+\nabla_i w)&\to-\frac{1}{\sqrt{h^{(0)}}}\epsilon_{ij}J^j_{(0)}\,,\qquad
J^j_{(0)}\to -\sqrt{h^{(0)}}\epsilon^{ij}(E_j+\nabla_j w)\,.
\end{align}
One can check that these transformations also leave \eqref{pos} invariant.

The S-duality invariance leads to consequences\footnote{An early paper that discussed the consequences of $S$-duality on conductivities
within holography and in a translationally invariant context is \cite{Hartnoll:2007ip}. A discussion of connections with results from magnetohydrodamics 
incorporating a phenomenological method of dissipating momentum appears in \cite{Hartnoll:2007ih}.}  for the DC conductivity.
Define $S_H\equiv (\rho_H,B_H,Z^{(0)},\vartheta^{(0)})$ to be the horizon data that is transformed under S-duality as in \eqref{stran}, 
and the {\it inverse} transformed quantities
$S'_H\equiv (B_H,-\rho_H,\frac{Z^{(0)}}{Z_{(0)}^2+\vartheta_{(0)}^2},-\frac{\vartheta^{(0)}}{Z_{(0)}^2+\vartheta_{(0)}^2})$. Then,
following the discussion in appendix E of \cite{Donos:2015bxe}, we deduce that
\begin{align}\label{dualtc}
\sigma^{ij}(S'_H)&=-\epsilon(ik)\sigma^{-1}_{kl}(S_H)\epsilon(lj)\,,\qquad
\alpha^{ij}(S'_H)=-\epsilon(ik)[\sigma^{-1}_{kl}\alpha^{lj}](S_H)\,,\cr
\bar{\alpha}^{ij}(S'_H)&=-[\bar{\alpha}^{ik}\sigma^{-1}_{kl}](S_H)\epsilon(lj)\,,\qquad
\bar{\kappa}^{ij}(S'_H)=\kappa^{ij}(S_H)\,,
\end{align}
where, as usual $\kappa^{ij}\equiv\bar{\kappa}^{ij}-T\bar{\alpha}^{ik}\sigma^{-1}_{kl}\alpha^{lj}$.
It is important to note that, in general, given one has a black hole solution with horizon data $S_H$, one is not guaranteed that there is in fact a black hole solution with horizon data $S_H'$, unless the equations of motion for the
full bulk field theory are themselves invariant under S-duality. Note that when the bulk field theory is invariant under
S-duality, then the transformations \eqref{sdualft} when evaluated on the horizon indeed give rise to \eqref{stran}.

Finally, let us consider a bulk $S$-duality invariant theory with $Z^2+\vartheta^2=1$ as in the consistent truncation of D=11 supergravity on an arbitrary seven dimensional Sasaki-Einstein space \cite{Gauntlett:2009bh}. A background solution of this theory that is symmetric under the combination of S-duality followed by time reversal has $\rho_H=B_H$ and we deduce, from \eqref{dualtc},\eqref{simpon}, that the conductivities must satisfy
\begin{align}\label{dualtctr}
\sigma^{ij}&=-\epsilon(jk)\sigma^{-1}_{kl}\epsilon(li)\,,\qquad
\alpha^{ij}=\bar{\alpha}^{jk}\sigma^{-1}_{kl}\epsilon(li)\,,\nn
\bar{\alpha}^{ij}&=\epsilon(jk)\sigma^{-1}_{kl}{\alpha}^{li}\,,\qquad\quad
\bar{\kappa}^{ij}=\bar{\kappa}^{ji}-T\bar{\alpha}^{jk}\sigma^{-1}_{kl}{\alpha}^{li}\,.
\end{align}
In particular, this implies that $\det(\sigma)=1$ and also $\det(\alpha)=\det(\bar{\alpha})$ (provided that the conductivities are finite).
This generalises analogous results for $\sigma^{ij}$ of \cite{Grozdanov:2015qia}, which studied the special case of Einstein-Maxwell theory and  solutions with $\rho_H=B_H=0$.

\section{Examples}\label{examples}

In this section we explore some of the effects of $S_\vartheta$ on the DC conductivity in various contexts.
In particular it naturally gives rise to an anomalous Hall conductivity, by which we mean a Hall current in the absence
of an applied external magnetic field. 

\subsection{Constant $\vartheta$ in the bulk}

The simplest case to consider, which has been already discussed in \cite{Iqbal:2008by} (see also \cite{Fischler:2015cma}), 
is when $\vartheta=\theta/(2\pi)^2$ is a constant in the bulk. In this
case $S_\vartheta$ is a total derivative and does not affect the bulk equations of motion. 
However, it does lead to a modification of the bulk currents so that at the horizon we have
\bea
J^i_{(0)}\to J^i_{(0)}-\frac{\theta}{(2\pi)^2}\epsilon(ij)(E_j+\partial_jw)\,,
\eea
with the extra term trivially conserved. This leads to an extra contribution to the Hall conductivity via,
$\sigma^{ij}\rightarrow\sigma^{ij}-\frac{\theta}{(2\pi)^2}\epsilon(ij)$,
with $\alpha,\bar\alpha$ and $\bar\kappa$ unchanged. 
Clearly constant $\vartheta$ gives rise to an anomalous Hall conductivity.

We also note that for this case the term $S_\vartheta$ give rise to a Chern-Simons term on the boundary. The $T$-duality transformation defined
by $\theta\to\theta+2\pi$ combines with $S$ duality transformation discussed earlier to generate $SL(2,Z)$ and this 
is discussed in \cite{Witten:2003ya} and also in \cite{Hartnoll:2007ip}.

It is also worth noting that for this case unless $\theta=0,\pi$ the theory breaks time reversal invariance (e.g. see \cite{Seiberg:2016rsg}).
In particular, the Onsager relations involve taking $\theta\to -\theta$.

\subsection{Non-constant $\vartheta$}
We now move on to more general setups in which $\vartheta$ is not a constant, but a function of pseudoscalar and scalar fields. 
Let us first consider holographic lattices in which the bulk gauge field
vanishes everywhere, $A_\mu=0$. In this case we see that there is a decoupling of $w,E_i$ from $v_i,p,\zeta_i$ in the Stokes equations.
In particular, to obtain the electrical conductivity matrix one should solve $\partial_i J^i_{(0)}=0$ for $w$. That is, on the horizon we need to solve
\begin{align}\label{divjays}
\nabla_i\Big(\big(Z^{(0)}h^{ij}_{(0)}-\vartheta^{(0)}\epsilon^{ij}\big)\left(E_j+{}{\nabla}_jw\right)\Big)=0\,.
\end{align}
Notice that the piece involving $\vartheta^{(0)}$ is not trivially conserved as it was when $\vartheta$ was constant.
Generically, solving \eqref{divjays} will give rise to a non-zero Hall response, with $\sigma^{xy}\ne 0$. 
We also
emphasise that if $\vartheta$ depends on a pseudoscalar that is dual to a relevant operator in the dual CFT, 
at the AdS boundary $\vartheta$ will vanish but, generically, $\vartheta^{(0)}$ will be non-zero; this underscores the role of the
black hole horizon in obtaining the DC Hall conductivity.

To make further explicit progress, still with $A_\mu=0$, we can consider the specific class of black hole backgrounds in which the
bulk fields only depend on the radial direction, with no dependence on the spatial coordinates\footnote{Note that a related
discussion is made in \cite{Iqbal:2008by} by imposing by hand the phenomenological condition $\vartheta=\vartheta(r)$. However, it was incorrectly concluded
that the Hall conductivity is associated with the value of $\vartheta(r)$ at $r\to\infty$, the AdS boundary.}. For
such backgrounds there is no dissipation of momentum and the thermal DC conductivity is necessarily infinite. However,
the electric conductivity is finite. Indeed, in this case all background quantities appearing
in \eqref{divjays} are constant and so we can solve the equation with $w=0$. This then leads to the finite DC conductivity matrix
\begin{align}\label{qchc}
\sigma^{ij}= \sqrt{h^{(0)}}\left(Z^{(0)}h^{ij}_{(0)}-\vartheta^{(0)}\epsilon^{ij} \right)\,.
\end{align}
Since this finite electrical conductivity is independent of any breaking of translation symmetry, in the context of a quantum critical theory, 
it is sometimes called the quantum critical conductivity. Here we see that $S_\vartheta$ leads to an anomalous Hall component.  

A simple way to obtain such black hole solutions is to consider metrics that are homogeneous and isotropic with respect to the spatial directions of the dual field theory
and just switch on a source for the pseudoscalar field $\phi(r)$ which is taken to be dual to a relevant operator in the dual field theory. 
In this type of setup, the source breaks the time reversal symmetry explicitly. A specific construction
of this type was made in section 8 of \cite{Gauntlett:2009bh}.

Interestingly, it is also possible to have constructions in which the quantum critical hall conductivity appears when the
time reversal symmetry is broken spontaneously and still\footnote{Note that in the interesting construction \cite{Andrade:2017leb} 
time reversal invariance is broken spontaneously but it uses a spatially modulated chemical potential. Without the latter, translation invariance is spontaneously broken \cite{Donos:2011bh} and the DC electrical conductivity is infinite.} 
with $A_\mu=0$. 
To see this, we first assume that the bulk theory of gravity depends on the pseudoscalar $\phi$ as
well as an additional scalar $\sigma$, which we take to be even under time reversal.  
The functions $V,Z,\vartheta$ will then depend on both $\phi$ and $\sigma$, with $V,Z$ even and $\vartheta$
odd under time reversal. It is not difficult to find choices for $V$ for
which there is a Poincar\'e invariant RG flow between the $AdS_4$ vacuum at infinity to another $AdS_4$ solution in the IR that
is driven by a deformation of the scalar field $\sigma(r)$ and for which $\phi=A_\mu=0$. In particular, one needs the scalar 
$\sigma$ to be dual to a relevant operator in the dual field theory and one imposes boundary conditions associated with sourcing this operator.
This source does not break the time reversal symmetry (nor the translation symmetry). 
We next demand that $V$ is chosen so that while $\phi$ is dual to a relevant 
operator in the UV $AdS_4$ solution, it violates the BF bound in the IR $AdS_4$ solution. This means that the IR vacuum is unstable and hence, with the same source for the relevant operator dual to 
$\sigma(r)$ still switched on in the UV, there will be a finite temperature phase transition at some critical temperature in
which a new branch of black holes appears with $\phi(r)\ne 0$. On this new branch
the operator dual to $\phi$ acquires an expectation value and thus the 
time reversal symmetry is broken 
spontaneously. In the black hole solutions describing this phase, one will find, generically, that the
pseudoscalar field $\phi(r)$ will be non-zero at the horizon, leading to non-vanishing $\vartheta^{(0)}$ and hence non-vanishing Hall conductivity.

An interesting aspect of the constructions we have just outlined, is that they provide a natural framework for
the temperature scaling of the Hall conductivity to be different to that of the longitudinal conductivity. Indeed these scalings will
be governed by the functions $Z^{(0)}$, $\vartheta^{(0)}$ appearing in \eqref{qchc}. It would be interesting to explore the range of different possibilities that are allowed taking into account the constraint on the bulk theories of gravity that would be imposed by
demanding that there are no further thermodynamic instabilities. These constructions could also be extended to include 
momentum dissipating effects in a natural way. For example, in the construction in which time reversal is broken spontaneously, one can add additional deformations in the UV which break translations without explicitly breaking time reversal invariance. One way to achieve this
is to have additional deformations of the scalar field $\sigma$ that depend on the spatial
coordinates of the dual field theory. One possibility would be a Q-lattice construction of the type discussed below.

One could also take such constructions and switch on a small magnetic field to explore the different scalings that are 
possible for the Hall angle. The term $S_\vartheta$ allows for more possibilities than those considered in 
\cite{Blake:2014yla}.

\subsection{Q-lattices}
The key idea of the Q-lattice construction \cite{Donos:2013eha} is to exploit a global symmetry in the bulk in order to obtain an ansatz for the bulk fields in which the dependence on the spatial directions is solved exactly. This leads to a system of ordinary differential equations for a set of functions or the bulk
fields that just depend on the holographic radial coordinate. It has been shown that the Stokes equations can be solved in terms of the horizon data explicitly for Q-lattices \cite{Donos:2015gia,Banks:2015wha,Donos:2015bxe}. 
This can easily be generalised when there is a $S_\vartheta$ term in the bulk theory of gravity. For example, suppose that
the pseudoscalar field is replaced with a complex scalar field. If $V$, $Z$ and $\vartheta$ only depend on the modulus of the complex pseudoscalar
then a Q-lattice construction can be developed by taking the phase to depend linearly on one of the spatial coordinates. 
In this case we will have $\vartheta$ to be independent of the spatial coordinates, $\partial_i\vartheta=0$.
We then find that the expressions for the conductivity are as in section 6.1 of \cite{Donos:2015bxe} with the modification
$\sigma^{ij}\rightarrow\sigma^{ij}-\vartheta^{(0)}\epsilon(ij)$. 

\subsection{One-dimensional lattices}
We now consider a class of black hole solutions in which the UV deformations 
break translations periodically in one of the two spatial directions, which we take to be $x$, with $x=x+L$.
We write the solution at the horizon\footnote{This ansatz can be justified by
noting that it is invariant under $y\to-y$, $t\to-t$, with the gauge-field going to minus itself,
which is a symmetry of the equations of motion.} as:
\begin{align}\label{2onedform}
h^{(0)}_{ij}\,dx^{i}dx^{j}&=\gamma(x) \,dx^{2}+\lambda(x)dy^2\,,\nn
F^{(0)}_{xy}&=B_H(x)\,,
\end{align}
with\footnote{Note that this corrects a sign typo in eq. (6.18) of \cite{Donos:2015bxe}.} $4\pi T\chi^{(0)}_i\equiv   (0, \chi(x))$,
$\phi^{(0)}=\phi^{(0)}(x)$ and $A^{(0)}_t=A^{(0)}_t(x)$. All functions\footnote{In particular, since $\chi(x)$ is periodic we are not allowing for NUT charges.} 
here are periodic in $x$ with period $L$.  The perturbing electric field source, $E_i$, and heat source, $\zeta_i$, are taken to be constant.

It is useful to define the following constant zero modes: 
\begin{align}
{B}=\int B_H,\qquad
{\rho}
=\int \rho_H,\qquad
{s}=
\int s_H\,,
\end{align}
where $\int$ refers to an integral averaged over a period of $x$, i.e. $\int\equiv (1/L)\int_0^L dx$, $\rho_H\equiv J^t=[(\gamma\lambda)^{1/2} Z^{(0)} A_t^{(0)}-\vartheta^{(0)} B_H]$ and
$s_H= (\gamma\lambda)^{1/2} (4\pi )$. 
We will also write
\begin{align}\label{pidf}
B_H=B +\partial_x \hat A_y\,,\qquad \rho_H=\rho +\partial_x C\,,
\end{align}
where $\hat A_y(x)$ and $C(x)$ are both periodic functions of $x$. 
Note that $B$ is the constant part of the magnetic field and $\rho$ is the constant total charge density of the dual field theory, neither
of which are renormalised in going to the UV. Furthermore, ${s}$ is the total entropy density of the dual field theory.

Here we consider three cases.
Firstly, when $B\ne 0$, second when $B=0$ but $B_H\ne 0$ and finally cases in which the gauge field vanishes everywhere.
In the appendix we have also included some special cases when the total zero mode of the charge density vanishes, $\rho=0$, which can be
achieve by imposing a symmetry on the class of solutions being considered.
The derivation for the conductivities follows that presented in \cite{Banks:2015wha,Donos:2015bxe} and so here we will just record the final
results. In order to do so we now define various quantities that appear in the final expressions.
We first define periodic functions $w_1(x)$ and $w_2(x)$ via:
\begin{align}
w_1(x)={\rho}\left(\frac{1}{{B}} \int^xB_H-\frac{1}{{\rho}}\int^x \rho_H \right)\, ,
\quad
w_2(x)=T{s}\left(\frac{1}{{B}}\int^x B_H- \frac{1}{{s}}\int^x s_H\right)\, ,
\end{align}
where $\int^x$ refers to an integral from some fiducial point $x=0$ to $x$ (with no division by $L$).
We next define the following five periodic functions
\begin{align}\label{youse}
u_1(x)&=\int^x\frac{\gamma^{1/2}\chi}{\lambda^{3/2}}-\frac{\int{}\frac{\gamma^{1/2}\chi}{\lambda^{3/2}}}{\int{}\frac{\gamma^{1/2}}{\lambda^{3/2}}}\int^x\frac{\gamma^{1/2}}{\lambda^{3/2}}\,,\nn
u_2(x)&=\int^x\frac{\gamma^{1/2}w_1}{\lambda^{3/2}}-\frac{\int{}\frac{\gamma^{1/2}w_1}{\lambda^{3/2}}}{\int{}\frac{\gamma^{1/2}}{\lambda^{3/2}}}\int^x\frac{\gamma^{1/2}}{\lambda^{3/2}}\,,\nn
u_3(x)&=\int^x\frac{\gamma^{1/2}w_2}{\lambda^{3/2}}-\frac{\int{}\frac{\gamma^{1/2}w_2}{\lambda^{3/2}}}{\int{}\frac{\gamma^{1/2}}{\lambda^{3/2}}}\int^x\frac{\gamma^{1/2}}{\lambda^{3/2}}\,,\nn
u_4(x)&=\int^x\frac{\gamma^{1/2}\hat A_y}{\lambda^{3/2}}-\frac{\int{}\frac{\gamma^{1/2}\hat A_y}{\lambda^{3/2}}}{\int{}\frac{\gamma^{1/2}}{\lambda^{3/2}}}\int^x\frac{\gamma^{1/2}}{\lambda^{3/2}}\,,\nn
u_5(x)&=\int^x\frac{\gamma^{1/2}C}{\lambda^{3/2}}-\frac{\int{}\frac{\gamma^{1/2}C}{\lambda^{3/2}}}{\int{}\frac{\gamma^{1/2}}{\lambda^{3/2}}}\int^x\frac{\gamma^{1/2}}{\lambda^{3/2}}\,.
\end{align}
These functions can be used to define the five by five matrix $\mathcal{U}$, with constant components given by
\begin{align}
\mathcal{U}_{ij}\equiv\int{}\frac{\lambda^{3/2}}{\gamma^{1/2}}\partial_xu_i\partial_xu_j\,.
\end{align}
Notice that
$\partial_x(\frac{\lambda^{3/2}}{\gamma^{1/2}}\partial_xu_i)$ has a simple form; for example
$\partial_x(\frac{\lambda^{3/2}}{\gamma^{1/2}}\partial_xu_1)=\partial_x\chi$.

\subsubsection{$B\ne0$: non-vanishing magnetic field zero mode}\label{bneqzero}
For this case, we obtain a finite response for general $E_i$, $\zeta_i$. Define the constant
\begin{align}\label{defex}
X=\int{}\frac{\left(\partial_{x} \lambda\right)^{2}}{\lambda^{5/2}\gamma^{1/2}}
+\int{}\frac{\left(\partial_{x}\phi^{(0)}\right)^{2}}{(\gamma\,\lambda)^{1/2}}\,+\int{}\frac{(\rho_H+B_H\vartheta^{(0)})^2}{\lambda Z^{(0)}(\gamma\lambda)^{1/2}}+\int{}\frac{B^2_HZ^{(0)}}{\lambda(\gamma\lambda)^{1/2}}+\mathcal{U}_{11}\,.
\end{align}
For the electric conductivity $\sigma$ we then find
\begin{align}
\sigma^{xx}&=0\,,\nn
\sigma^{yy}&=\mathcal{U}_{22}+\int{}\frac{\gamma^{1/2}Z^{(0)}}{\lambda^{1/2}}+\int{}(\frac{{\rho}}{{B}}+\vartheta^{(0)})^2\frac{\gamma^{1/2}}{\lambda^{1/2}Z^{(0)}}\nn
&\qquad\qquad
-\frac{1}{X}\left(
\mathcal{U}_{12}-\int{}(\frac{{\rho}}{{B}}+\vartheta^{(0)})\frac{(\rho_H+B_H\vartheta^{(0)})}{\lambda Z^{(0)}}-\int{}\frac{B_HZ^{(0)}}{\lambda}\right)^2\,,\nn
\sigma^{xy}&=-\sigma^{yx}=\frac{{\rho}}{{B}}\,.
\end{align}
For the thermoelectric conductivities $\alpha$ and $\bar\alpha$ we find
\begin{align}
\alpha^{xx}&=\bar{\alpha}^{xx}=0\,,\nn
\alpha^{yy}&=\bar{\alpha}^{yy}=\frac{\mathcal{U}_{23}}{T}+\frac{{s}}{{B}}\int{}(\frac{{\rho}}{{B}}+\vartheta^{(0)})\frac{\gamma^{1/2}}{\lambda^{1/2}Z^{(0)}}\nn
-&\frac{1}{X}\Big(\mathcal{U}_{12}-\int{}(\frac{{\rho}}{{B}}+\vartheta^{(0)})\frac{(\rho_H+B_H\vartheta^{(0)})}{\lambda Z^{(0)}} -\int{}\frac{B_HZ^{(0)}}{\lambda}\Big)\Big(\frac{\mathcal{U}_{13}}{T}-\frac{{s}}{{B}} \int{}\frac{(\rho_H+B_H\vartheta^{(0)})}{\lambda Z^{(0)}}\Big)\,,\nn
\alpha^{xy}&=-\bar{\alpha}^{yx}=\frac{{s}}{{B}}\,,\nn
\alpha^{yx}&=-\bar{\alpha}^{xy}=\frac{4\pi}{X}\Big(\mathcal{U}_{12}-\int{}(\frac{{\rho}}{{B}}+\vartheta^{(0)})\frac{(\rho_H+B_H\vartheta^{(0)})}{\lambda Z^{(0)}} -\int{}\frac{B_HZ^{(0)}}{\lambda}\Big)\,.
\end{align}
Finally, for the thermal conductivity $\bar\kappa$ we have
\begin{align}\bar{\kappa}^{xx}&=\frac{16\pi^2 T}{X}\,,\nn
\bar{\kappa}^{yy}&=\frac{\mathcal{U}_{33}}{T}+\frac{{s}^2T}{{B}^2}\int{}\frac{\gamma^{1/2}}{\lambda^{1/2}Z^{(0)}}-\frac{T}{X}\Big(\frac{\mathcal{U}_{13}}{T}-\frac{{s}}{{B}}\int{}\frac{(\rho_H+B_H\vartheta^{(0)})}{\lambda Z^{(0)}}\Big)^2\,,\nn
\bar{\kappa}^{xy}&=-\bar{\kappa}^{yx}=-\frac{4\pi T}{X}\Big(\frac{\mathcal{U}_{13}}{T}-\frac{{s}}{{B}} \int{}\frac{(\rho_H+B_H\vartheta^{(0)})}{\lambda Z^{(0)}}\Big)\,.
\end{align}
If we set $\vartheta^{(0)}=0$ then we recover the expressions given in \cite{Donos:2015bxe}.
We have not managed to obtain significantly simpler expressions for the inverse of the conductivity matrix, which one might have hoped for
following a discussion in section 5.10.1 of \cite{Hartnoll:2016apf}.

\subsubsection{$B$=0: vanishing magnetic field zero mode}
We next consider the case with $B=0$ but still allowing for the possibility of $B_H\ne 0$. In other words there can be magnetisation currents at the horizon of the form
$B_H=\partial_x \hat A_y$ with $\hat A_y$ a periodic function. This case can arise when the time-reversal invariance is explicitly broken: for example it will occur 
when there is a source for the gauge field in the holographic lattice 
leading to pure magnetisation currents in the dual field theory with $B=0$ (i.e. $A^{(\infty)}_i \ne 0$ in \eqref{asmet} is a periodic function). 
It also covers the constructions in \cite{Andrade:2017leb}
which the time-reversal invariance is broken spontaneously with $A^{(\infty)}_i =\phi^{(\infty)}=g_{ti}^{(\infty)}=0$, but nevertheless $B_H\ne 0$.

Generically, in order to find solutions to the Stokes equations, associated with a
finite DC response in the dual field theory, we can have $E_x,\zeta_x\ne0$ but should set
\begin{align}\label{settozero}
E_y=\zeta_y=0\,.
\end{align}
We now define
\bea
\tilde{X}=&X(\int{}\frac{\gamma^{1/2}}{\lambda^{1/2}Z^{(0)}}+\mathcal{U}_{44})-(\int{}\frac{(\rho_H+B_H\vartheta^{(0)})}{\lambda Z^{(0)}}-\mathcal{U}_{14} )^2\,.
\eea
Calculating the electric and thermal currents in the $x$ direction we find:
\begin{align}\label{xdcnbz}
\sigma^{xx}&=\frac{X}{\tilde{X}}\,,\nn
\alpha^{xx}=\bar{\alpha}^{xx}&=\frac{4\pi}{\tilde{X}}  \Big(\int{}\frac{(\rho_H+B_H\vartheta^{(0)})}{\lambda Z^{(0)}}-\mathcal{U}_{14} \Big)\,,\nn
\bar{\kappa}^{xx}&=\frac{(4\pi)^2T}{\tilde{X}} \Big(\int{}\frac{\gamma^{1/2}}{\lambda^{1/2}Z^{(0)}}+\mathcal{U}_{44}\Big)\,.
\end{align}
Setting $\vartheta^{(0)}=B_H=\chi=0$ in these expressions leads to the results given in \cite{Banks:2015wha} (with $X_{there}$ identified with 
$\tilde X$).
We also observe that $T\sigma^{xx} \bar{\kappa}^{xx}-(T\alpha^{xx})^2=(4\pi T)^2/\tilde X$ and hence we can also write simple expressions for the inverse of the conductivities, in this two times two block, similar to \cite{Hartnoll:2016apf}. Furthermore,
$T\kappa^{xx}\equiv T\bar{\kappa}^{xx}-(T\alpha^{xx})^2/\sigma^{xx}=(4\pi T)^2/ X$.

Note that, in general, when $B=0$ and in addition $\rho\ne0$, the Stokes equations do not lead to a unique answer for the electric and thermal currents in the $y$ direction due to the existence of an undetermined integration constant; this zero mode arises because of the translation
invariance in the $y$ direction. This lack of uniqueness corresponds to the appearance of delta functions in the AC conductivities associated with these currents, at zero frequency. This is also related to the necessity of \eqref{settozero} to obtain solutions.

\subsubsection{Vanishing gauge-fields}\label{vangf}

In the previous sub-section with $B=0$, if we also set $\rho=0$ then we can allow for $E_y\ne 0$ and, in addition, we find a finite response
for the DC electric current in the $y$ direction. 
We still must impose $\zeta_y=0$ to find solutions and, furthermore, we still cannot obtain the DC heat current in the $y$ direction from the Stokes equations, again
due to the presence of a zero mode associated with an infinite DC response.
General expressions can be written down for the finite conductivities but we will omit them both because they are rather long
and, in addition, the condition $\rho=0$ will occur, generically, only if there is some symmetry principle protecting it.
Below we will write down the expressions when the gauge-fields vanish, which obviously has $\rho=0$, and in the appendix we
have recorded the results for some other special cases.

Specifically, for vanishing gauge fields, we allow $E_x,E_y,\zeta_x\ne0$ but in order to find solutions we should still set $\zeta_y=0$.
The finite thermoelectric conductivities can then be written in the simple form
 \begin{align}
 \sigma^{xx}&=\Big(\int{}\frac{\gamma^{1/2}}{\lambda^{1/2}Z^{(0)}}\Big)^{-1}\,,\nn
  \sigma^{yy}&=\int{}\frac{\gamma^{1/2}[(Z^{(0)})^2+(\vartheta^{(0)})^2   ]  }{\lambda^{1/2}Z^{(0)}}  -\Big(\int{}\frac{\gamma^{1/2}}{\lambda^{1/2}Z^{(0)}}\Big)^{-1}\Big(\int{}\frac{\gamma^{1/2}\vartheta^{(0)}}{\lambda^{1/2}Z^{(0)}}\Big)^2\,,\cr
 \sigma^{xy}&=- \sigma^{yx}=-\Big(\int{}\frac{\gamma^{1/2}}{\lambda^{1/2}Z^{(0)}}\Big)^{-1}\int{}\frac{\gamma^{1/2}\vartheta^{(0)}}{\lambda^{1/2}Z^{(0)}}\,,\cr
\alpha^{xx}&=\bar{\alpha}^{xx}= \bar{\alpha}^{xy}=0\,,\cr
\bar{\kappa}^{xx}&=\frac{(4\pi)^2T}{{X}} \,.
 \end{align}
 where $X$ now has the simpler form $X=\int{}\frac{\left(\partial_{x} \lambda\right)^{2}}{\lambda^{5/2}\gamma^{1/2}}
+\int{}\frac{\left(\partial_{x}\phi^{(0)}\right)^{2}}{(\gamma\,\lambda)^{1/2}}\,+\mathcal{U}_{11}$. Note that, consistent
with the discussion at the end of section \ref{sdualsec}, if $(Z^{(0)})^2+(\vartheta^{(0)})^2=1$ then $\det\sigma=1$.
We also note that to have non-vanishing Hall currents for this set up, requires that
$\vartheta^{(0)}\ne 0$.

\section{Final Comments}

In this paper we extended our understanding of thermoelectric DC conductivity within the context of holography using the approach developed
in \cite{Donos:2015gia,Banks:2015wha,Donos:2015bxe,Donos:2017oym}.
We showed that the Stokes equations for the auxiliary fluid living on the black hole horizon can 
be derived by varying a functional that depends on both the fluid variables and the black hole geometry of
interest as well as the time reversed quantities. This then leads to a simple
derivation of the Onsager relations for the DC conductivity.

We also generalised the formalism to include the term
$S_\vartheta\sim\int \vartheta F\wedge F$ in the bulk action in four spacetime dimensions. 
We outlined several different constructions
and discussed the associated anomalous Hall conductivity. We also presented the solution of the Stokes equations for 
the case of holographic lattices that only depend on one of the spatial coordinates.
For holographic lattices of pure Einstein gravity in the hydrodynamic limit, i.e. in the limit in which the temperature is the highest scale, it has been
shown how the auxiliary fluid on the black hole horizon can be identified with the physical fluid of the dual CFT
\cite{Banks:2016krz} in the presence of external DC sources. It would be interesting to extend this analysis to also incorporate scalar fields and gauge fields, including
the $S_\vartheta$ term. We anticipate that the Stokes equations that we have derived in this paper could also be obtained from
a suitable generalisation of the parity violating relativistic hydrodynamics of \cite{Jensen:2011xb}
to include additional scalar fields.

It would also be interesting to include the effects of a $S_{\hat\vartheta}\sim\int \hat \vartheta R\wedge R$ term in the action where $\hat\vartheta$ is a function of the dynamical fields. Such terms have been
considered in various hydrodynamic contexts in holography including \cite{Saremi:2011ab,Delsate:2011qp,Kimura:2011ef,Chen:2011fs}.
The currents and the Stokes equations relevant for obtaining the DC conductivity can be obtained by utilising the tools developed in \cite{Donos:2017oym} to study theories with higher derivatives.

\section*{Acknowledgements}
We thank Barak Kol for a correspondence.
The work of JPG and LM is supported by the European Research Council under the European Union's Seventh Framework Programme (FP7/2007-2013), ERC Grant agreement ADG 339140. The work of JPG and TG is supported by STFC grant ST/L00044X/1. JPG is also
supported by EPSRC grant EP/K034456/1 and supported as a KIAS Scholar and as a Visiting Fellow at the Perimeter Institute.

\appendix
\section{One-dimensional lattices with $\rho=0$}

Here we record the DC conductivities for some additional one-dimensional lattices that have vanishing total charge density, $\rho=0$, arising
because of
some underlying symmetry. In order to find solutions to the Stokes equations, generically we will take $E_x,E_y,\zeta_x\ne0$ and
$\zeta_y=0$, and we also note that $Q^y_{(0)}$ is not uniquely determined, but for
cases when we have $B\ne 0$ we will also be able to allow for $\zeta_y\ne 0$ and find a unique $Q^y_{(0)}$. 
It is worth noting that when $\rho=0$ we see from
\eqref{pidf} and \eqref{youse} that we have $u_2(x)=-u_5(x)$.

\subsection{$\rho=0$ and $A_t\ne 0$}
We discuss two cases in which we can have vanishing total charge density, $\rho=0$, and also have 
$A_t\ne 0$.

The first case exploits the fact that the action \eqref{actover} 
is invariant under $x\rightarrow-x$, $t\rightarrow-t$, leaving the fields invariant\footnote{A very similar result to that which we describe here also occurs for actions in which $x\rightarrow-x$, $t\rightarrow-t$ combined with $\phi\to -\phi$ is a symmetry, which arises when $V, Z$ and also $\vartheta$ are even functions of $\phi$.}.
Taking a one-dimensional ansatz that is invariant under this symmetry implies that in the horizon quantities 
\eqref{2onedform} we have that $A^{(0)}_t$ and $\chi^{(0)}_y$ must be odd functions of $x$ while $A^{(0)}_y$ and $h^{(0)}_{ij}$ must be even functions of $x$. In addition, $\phi^{(0)}$ is an even function of $x$ and hence
$Z, V$ and $\vartheta$ will all be even functions of $x$.  Notice that since $B_H$ is an odd and periodic function of $x$ we must have vanishing zero mode for the magnetic field, $B=0$. In addition, we also have, 
$\mathcal{U}_{12}=\mathcal{U}_{14}=\mathcal{U}_{15}=0$.

For this case we find that the finite DC conductivities are given by
 \bea
 \sigma^{xx}&=&(\int\frac{\gamma^{1/2}}{\lambda^{1/2}Z^{(0)}}+\mathcal{U}_{44})^{-1}\,,\cr\cr
 \sigma^{yy}&=&\int\frac{\gamma^{1/2}Z^{(0)}}{\lambda^{1/2}}+\int\frac{\gamma^{1/2}(\vartheta^{(0)})^2}{\lambda^{1/2}Z^{(0)}}+\mathcal{U}_{55}-(\int\frac{\gamma^{1/2}\vartheta^{(0)}}{\lambda^{1/2}Z^{(0)}}-\mathcal{U}_{45})^2(\int\frac{\gamma^{1/2}}{\lambda^{1/2}Z^{(0)}}+\mathcal{U}_{44})^{-1}\,,\cr\cr
 \bar{\kappa}^{xx}&=&\frac{(4\pi)^2T}{X}\,, \cr
 \sigma^{xy}&=&- \sigma^{yx}=-(\int\frac{\gamma^{1/2}\vartheta^{(0)}}{\lambda^{1/2}Z^{(0)}}-\mathcal{U}_{45})(\int\frac{\gamma^{1/2}}{\lambda^{1/2}Z^{(0)}}+\mathcal{U}_{44})^{-1}\,,\cr\cr
  \bar{\alpha}^{xy}&=&-\alpha^{yx}=\alpha^{xx}=\bar{\alpha}^{xx}=0\,,
 \eea
where $X$ is as in \eqref{defex}. We note that we have a non-zero Hall conductivity if $\vartheta^{(0)}\ne 0$ or due to
the non-vanishing of $\mathcal{U}_{45}$, which derives from local charge density and local magnetisation currents. It would be
interesting to find a physical realisation of this novel mechanism.

The second case arises when the action \eqref{actover} is symmetric under\footnote{We can also do something similar for actions invariant under 
$x\rightarrow-x$, $A\rightarrow-A$, $\phi\rightarrow\phi$, which requires $\vartheta=0$.}
 $x\rightarrow-x$, $A\rightarrow-A$, $\phi\rightarrow-\phi$.
This is a symmetry of the action provided that $\vartheta$ is an odd function of $\phi$ while $V, Z$ are even functions of $\phi$.
Taking an ansatz
that is invariant under this symmetry implies that in the horizon quantities 
\eqref{2onedform} we have that  $A^{(0)}_t$, $A^{(0)}_y$ must be odd functions of $x$ while $\chi^{(0)}_y$ and $h^{(0)}_{ij}$ must be even functions of $x$. In addition, $\phi^{(0)}$ must be an odd function of $x$ and hence $Z, V$ will be even functions and $\vartheta$ will be an odd function of $x$.  
One can check that we have $\mathcal{U}_{13}=\mathcal{U}_{14}=\mathcal{U}_{23}=\mathcal{U}_{24}=\mathcal{U}_{35}=\mathcal{U}_{45}=0$.

For vanishing zero mode for the magnetic field, $B=0$, the finite DC conductivities are given by
 \bea\label{aytwo}
 \sigma^{xx}&=&(\int\frac{\gamma^{1/2}}{\lambda^{1/2}Z^{(0)}}+\mathcal{U}_{44})^{-1}\,,\cr\cr
 \sigma^{yy}&=&\int\frac{\gamma^{1/2}Z^{(0)}}{\lambda^{1/2}}+\int\frac{\gamma^{1/2}(\vartheta^{(0)})^2}{\lambda^{1/2}Z^{(0)}}+\mathcal{U}_{55}-\frac{1}{X}(\int\frac{B_HZ^{(0)}}{\lambda}+\int\frac{(\rho_H+\vartheta^{(0)}B_H)}{\lambda Z^{(0)}}\vartheta^{(0)}+\mathcal{U}_{15})^2\,,\cr\cr
 \bar{\kappa}^{xx}&=&\frac{(4\pi)^2T}{X}\,, \cr\cr
 \bar{\alpha}^{xy}&=&-\alpha^{yx}=\frac{4\pi}{X}(\int\frac{B_HZ^{(0)}}{ \lambda}+\int\frac{(\rho_H+\vartheta^{(0)}B_H)}{\lambda Z^{(0)}}\vartheta^{(0)}+\mathcal{U}_{15})\,,\cr\cr
  \sigma^{xy}&=&- \sigma^{yx}= \alpha^{xx}=\bar{\alpha}^{xx}=0\,,
 \eea
where $X$ is as in \eqref{defex}.
It is interesting to highlight the three contributions to $\alpha^{yx}$ and $\bar{\alpha}^{xy}$. Let us discuss $\alpha^{yx}$ which
is the statement that an electric current in the $y$ direction is caused by a thermal gradient in the $x$ direction. 
The first term on the right hand side does not involve the charge density but just the local $B_H$ and is associated with
particle hole pairs moving in opposite directions due to $B_H$.  The second term involves $\vartheta^{(0)}$. The final term is associated 
with a particularly novel effect arising from a combination of local charge density and the local function $\chi^{(0)}_y$.
For this case we can also allow for non-vanishing zero mode of the magnetic field $B\ne0$ and also $\zeta_y\ne 0$. The resulting
conductivities can be obtained directly from the expressions given in section \ref{bneqzero} after setting $\rho=\mathcal{U}_{23}
=\mathcal{U}_{13}=0$ as well as noting that some other integrals vanish due to the fact that they
are  the averaged integral of an
odd and periodic function of $x$. 

Finally, we note that a one-dimensional holographic ionic lattice based on a single wave mode and vanishing chemical potential, 
as well as $\chi=A_y=0$ was studied in \cite{Chesler:2013qla}. Specifically, the deformation that was considered had 
$A^{(\infty)}_t=\mu\cos kx$. After making the shift $x\to x-(\pi/2k)$ we get $A^{(\infty)}_t=\mu\sin kx$, which is an odd function of $x$.
This example therefore fits into both of the classes above. In other words, it is possible to generalise the ansatz for the 
one-dimensional holographic lattice of \cite{Chesler:2013qla} in two different ways, while maintaining $\rho=0$.

\subsection{A case with $\vartheta=0$ and $A_t=0$}
When $\vartheta=0$ the action \eqref{actover} is invariant under the symmetry $t\to -t$. By
restricting to configurations that are invariant under this symmetry implies, in particular, $A_t=g_{ty}=0$. Notice that for this case we not only have $\rho=0$ but we have $\rho_H=0$.
We also have the components $\mathcal{U}_{ij}$ vanish if $i$ or $j$ are either 1,2 or 5.
For configurations with vanishing zero mode for the magnetic field, $B=0$, we find that the finite DC conductivity matrix elements are 
\bea
 \sigma^{xx}&=&(\int\frac{\gamma^{1/2}}{\lambda^{1/2}Z^{(0)}}+\mathcal{U}_{44})^{-1}\,,\cr\cr
 \sigma^{yy}&=&\int\frac{\gamma^{1/2}Z^{(0)}}{\lambda^{1/2}}-\frac{1}{X}(\int\frac{B_HZ^{(0)}}{\lambda})^2\,,\cr\cr
 \bar{\kappa}^{xx}&=&(4\pi)^2\frac{T}{X}\,,\cr\cr
  \bar{\alpha}^{xy}&=&-\alpha^{yx}=\frac{4\pi}{X}(\int\frac{B_HZ^{(0)}}{ \lambda})\,,\cr\cr
 \sigma^{xy}&=&- \sigma^{yx}= \alpha^{xx}=\bar{\alpha}^{xx}=0
 \eea
 where $X=\int \frac{\left(\partial_{x} \lambda\right)^{2}}{\lambda^{5/2}\gamma^{1/2}}
+\int\frac{\left(\partial_{x}\phi^{(0)}\right)^{2}}{(\gamma\,\lambda)^{1/2}}\,+\int\frac{B_H^2Z^{(0)}}{\lambda(\gamma\lambda)^{1/2}}$.
The non-vanishing of $\alpha^{yx}$ and $\bar{\alpha}^{xy}$ is coming from $B_H\ne0$ as discussed below \eqref{aytwo}.
For this case we can also allow for non-vanishing zero mode of the magnetic field $B\ne0$ and also $\zeta_y\ne 0$. The resulting
conductivities can be obtained from the expressions given in section \ref{bneqzero} after setting
$\rho_H=\vartheta=\mathcal{U}_{12}=\mathcal{U}_{13}=\mathcal{U}_{23}=0$.

\providecommand{\href}[2]{#2}\begingroup\raggedright\endgroup

\end{document}